# A Laterally Vibrating Lithium Niobate MEMS Resonator Array Operating at 500°C in Air

Savannah R. Benbrook, Caitlin A. Chapin, Ruochen Lu, Yansong Yang, Songbin Gong, *Senior Member, IEEE*, and Debbie G. Senesky, *Member, IEEE*

*Abstract*—This paper is the first report of the high-temperature characteristics of a laterally vibrating piezoelectric lithium niobate (LiNbO$_3$) MEMS resonator array up to 500°C in air. After a high-temperature burn-in treatment, device quality factor (Q) is enhanced to 508 and the resonance shifts to a lower frequency and remains stable up to 500°C. During subsequent in situ high-temperature testing, the resonant frequencies of two coupled shear horizontal (SH0) modes in the array are 87.36 MHz and 87.21 MHz at 25°C and 84.56 MHz and 84.39 MHz at 500°C, correspondingly, representing a -3% shift in frequency over the temperature range. Upon cooling to room temperature, the resonant frequency returns to 87.36 MHz, demonstrating recoverability of device performance. The first- and second-order temperature coefficient of frequency (TCF) are found to be -95.27 ppm/°C and 57.5 ppb/°C$^2$ for resonant mode A, and -95.43 ppm/°C and 55.8 ppb/°C$^2$ for resonant mode B, respectively. The temperature-dependent quality factor (Q) and electromechanical coupling coefficient ($k_t^2$) are extracted and reported. Device Q decreases to 334 after high-temperature exposure, while $k_t^2$ increases to 12.40%. This work supports the use of piezoelectric LiNbO$_3$ as a material platform for harsh environment radio-frequency (RF) resonant sensors (e.g. temperature and infrared).

*Index Terms*—Lithium Niobate, RF MEMS, Piezoelectric Resonators, High-Temperature, SH0 mode

## I. Introduction

RADIO FREQUENCY (RF) components capable of operating within environments extreme in temperature, pressure, corrosion, and radiation are desirable in a variety of industries including aerospace, military, automotive, and energy harvesting. For instance, reliable devices that can combine passive RF/wireless signal processing with sensing modalities are particularly crucial to the longevity and scope of many proposed space missions, such as exploration of hostile planets such as Venus (surface temperature 465°C). Such

This article was submitted December 9, 2019. This work was supported in part by the National Aeronautics and Space Agency through the High Operating Temperature Technology program under grant number NNX17AG44G.
S. R. Benbrook, C. A. Chapin, and D. G. Senesky are affiliated with the Department of Aeronautics and Astronautics at Stanford University, Stanford, CA 94305 USA (e-mail: sbenbroo@stanford.edu, dsenesky@stanford.edu).
R. Lu, Y. Yang, and S. Gong are with the Department of Electrical and Computer Engineering, University of Illinois at Urbana-Champaign, Urbana, IL 61801 USA (email: songbin@illinois.edu).

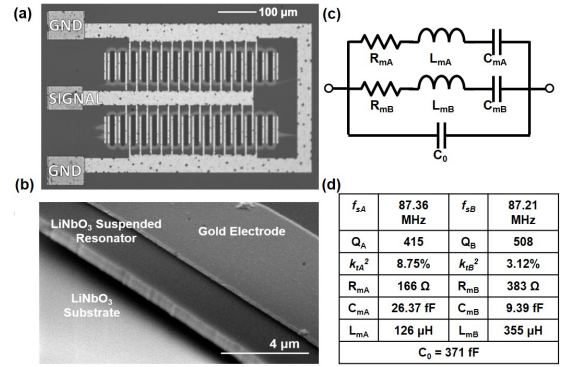

Fig. 1. Optical image (top-view) of the microfabricated LiNbO$_3$ MEMS resonator array post burn-in (a), SEM image of an individual resonator in the array (b), modified Butterworth-Van Dyke (MBVD) model for the array at 25°C after burn-in (c) and summary of key extracted device parameters from the MBDV model (d).

devices could potentially eliminate the need for active electronics for in-sensor signal processing and seamlessly integrate sensing and wireless readout in one battery-less package. To this end, piezoelectric RF microelectromechanical systems (MEMS) devices are particularly promising as they offer compact and uncooled sensing solutions that can survive on hot planets [1]–[4], as well as other Earth-based systems (e.g. oil and gas, geothermal).

Several piezoelectric devices made with materials such as langasite (La$_3$Ga$_5$SiO$_{14}$), aluminum nitride (AlN), and gallium nitride (GaN) have been examined for high-temperature RF applications [5]–[12]. Yet, the effective electromechanical coupling of the devices reported in these demonstrations is limited ($k_t^2$ < 1.4%) due to inherently low piezoelectric coupling coefficients ($K^2$) in the selected materials. The performance of piezoelectric lithium niobate (LiNbO$_3$) surface acoustic wave (SAW) devices has been investigated at elevated temperatures as high as 600°C in air, due to the attractive material properties of stoichiometric LiNbO$_3$ like a high Curie point of 1140°C, chemical inertness, and ability to maintain its favorable piezoelectric properties up to 750°C [13]–[18]. However, the performance of LiNbO$_3$ SAW devices is still limited by only moderate coupling ($k_t^2 \approx 1.5\%$) and energy leakage into the substrate. The low coupling and high loss in aforementioned platforms fundamentally limit the bandwidth, signal-to-noise ratio, and multiplexing capabilities needed for effective and robust passive wireless sensing.

Therefore, to address the need for low-loss, high coupling, and harsh environment capable wireless sensors, we examine the use of $LiNbO_3$ laterally vibrating resonators (LVRs). A distinct advantage of the $LiNbO_3$ LVR is its high electromechanical coupling coefficient ($k_t^2$ up to 30%) [19]. High $k_t^2$ values, along with comparable quality factors to other materials, can result in higher figure of merit (FOM = $k_t^2$ x Q), lower insertion loss, and wider bandwidths in $LiNbO_3$ LVRs. In this letter, we present the first investigation on the high-temperature characteristics of an array of $LiNbO_3$ laterally vibrating MEMS resonators ($LiNbO_3$ LVRs) up to 500°C in air. The laterally vibrating MEMS device exhibits recoverability after 500°C exposure, a high temperature coefficient of frequency (TCF), and high $k_t^2$, demonstrating the potential of this materials platform for low-loss and ultra-sensitive extreme environment RF sensing.

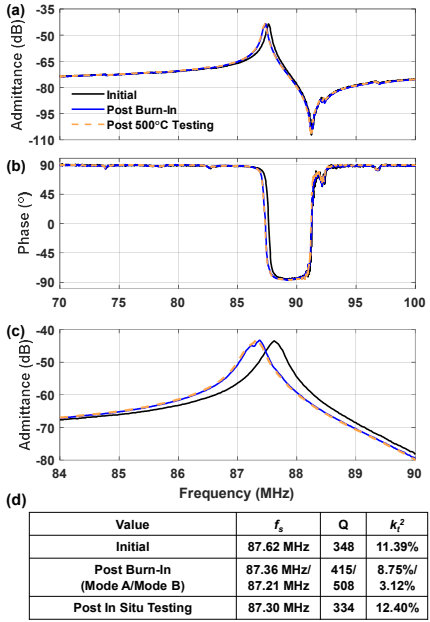

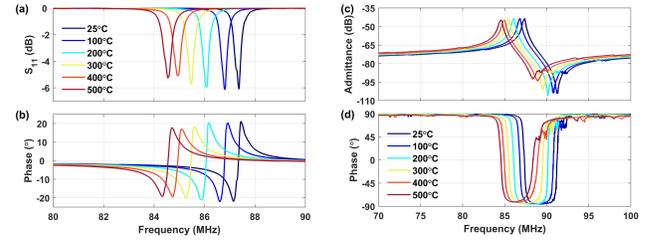

Fig. 3. Measured amplitude (a) and phase (b) of the $S_{11}$ parameters and measured amplitude (c) and phase (d) of the admittance of the $LiNbO_3$ MEMS resonator array from 25°C to 500°C.

| Value | $f_s$ | Q | $k_t^2$ |
|---|---|---|---|
| Initial | 87.62 MHz | 348 | 11.39% |
| Post Burn-In (Mode A/Mode B) | 87.36 MHz/ 87.21 MHz | 415/ 508 | 8.75%/ 3.12% |
| Post In Situ Testing | 87.30 MHz | 334 | 12.40% |

Fig. 2. Measured amplitude (a) and phase (b) of the admittance of the $LiNbO_3$ MEMS resonator array at 25°C initially, post burn-in, and post in-situ testing up to 500°C and back, with close-up of the resonance (c). Summary of extracted key parameters shown in (d).

## II. DEVICE DESIGN AND EXPERIMENTAL SETUP

Optical and scanning electron microscopy (SEM) images of the microfabricated $LiNbO_3$ MEMS resonator array are shown in Fig. 1a and 1b, respectively. The one-port array consists of 26 identical laterally vibrating resonators (LVRs). Each individual resonator has two 100-nm-thick gold interdigitated electrodes acting as signal and ground on 700-nm-thick suspended stoichiometric $LiNbO_3$. The resonator array was fabricated on an X-cut $LiNbO_3$ epitaxial thin film on nonstoichiometric $LiNbO_3$ substrate at -10° to the Y-axis to excite the shear horizontal mode (SH0). The total device dimensions are 728 μm by 440 μm. Further fabrication and design details can be found in prior work [20], [21]. The modified Butterworth–Van Dyke (MBVD) model obtained from the 25°C post burn-in device admittance measurements is shown in Fig. 1c. The MBVD model has two motional branches that represent two coupled SH0 modes very close to each other in resonance [22]. The existence of two modes is due to slight variation in mechanical boundary conditions, and thus resonant frequencies, among the individual resonators in the array. The extracted parameters of both modes, denoted Mode A ($f_{sA}$, $k_{tA}^2$, $Q_A$) and Mode B ($f_{sB}$, $k_{tB}^2$, $Q_B$), are shown in Fig. 1d.

The experimental setup consisted of a network analyzer (Agilent E5063A) connected by 50 Ω impendence coaxial cable to a custom high-temperature ground-signal-ground (GSG) probe (Supplier: GGB Industries). As a further precaution against detrimental heating effects on the system, aluminized heat shielding was affixed to the bottom of the probe body and wrapped around the coaxial cable region closest to the probe. The probe is housed within an open-air high-temperature probe station (Supplier: Signatone Inc.) equipped with a proportional-integral-derivative (PID) controlled chuck capable of heating up to 600°C in air.

Using this setup, the resonator array was first subjected to a burn-in cycle up to 500°C and back to room temperature in air. The purpose of the burn-in treatment was to stabilize any annealing effects (e.g., alloying due to diffusion of metal contacts) in the device. Initial and post-burn-in data was acquired. The device performance was then characterized in situ from 25°C to 500°C in 25°C increments. To allow for temperature stabilization, the device was held at each temperature for twenty minutes before data was acquired. After cooling back down from 500°C to 25°C, the device was measured again (post in situ).

## III. RESULTS AND DISCUSSION

Fig. 2 shows the measured admittance ($Y_{11}$) of the $LiNbO_3$ MEMS resonator array initially, post burn-in, and post in-situ testing. Extracted device parameters are listed in Fig. 2d. During initial and post in situ testing device measurements, only one SH0 mode is found and extracted. After the burn-in temperature cycle, the resonance peak has shifted down in frequency. The peak does not experience any further significant shift after the in-situ testing up to 500°C and back (Fig. 2a and 2c). This indicates that the burn-in temperature cycle stabilizes the resonant frequency, likely due to an annealing effect on the gold electrodes. While the burn-in procedure enhanced Q, subsequent in situ high-temperature testing reduced Q below its initial value. The total electromechanical coupled, $k_t^2$, of the array increases by a small amount from initial to post in situ device testing. The changes in Q and $k_t^2$ are likely caused by changes in the residual stress introduced during the fabrication process.

The $S_{11}$ parameters of the $LiNbO_3$ MEMS resonator array measured in situ during temperature ramp up from 25°C to 500°C are normalized to 50 Ω impedance and shown in Fig. 3a and 3b. The admittance ($Y_{11}$) parameters over the temperature range are shown in Fig. 3c and 3d. Both the magnitude and phase of the admittance experience a downward shift in frequency from 25°C to 500°C as expected, due to material softening.

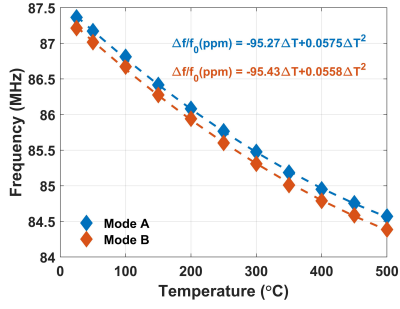

Fig. 4. Measured temperature dependence of the shear horizontal (SH0) mode resonant frequency LiNbO$_3$ MEMS resonator array up to 500°C.

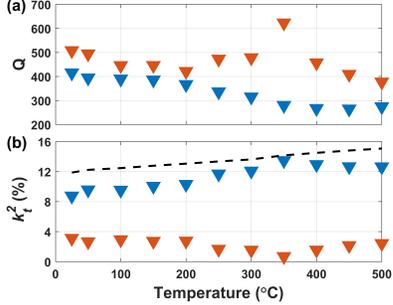

Fig. 5. Extracted Q (a) and $k_t^2$ (%) (b) as a function of temperature. Mode A and B are represented by blue and orange triangles, respectively. The sum of $k_t^2$ of both modes is represented by the dashed black line.

Fig. 4 shows the temperature-dependent resonant frequency over the temperature range for both coupled SH0 modes A and B. The resonance of device mode A is 87.36 MHz at 25°C and 84.56 MHz at 500°C, or a -3.2% shift. The resonance of device mode B is 87.21 MHz at 25°C and 84.39 MHz at 500°C. Upon cooling to room temperature, the resonant frequency returned to 87.30 MHz and only one mode is present, demonstrating recoverability of device performance after the twenty minute high-temperature exposure. Experimental frequency versus temperature results are best fit with quadratic equations as expected and are displayed on Fig. 4 as dashed lines. Using 25°C as the reference temperature, the fractional frequency variation for mode A is found

$$\frac{\Delta f}{f_0} = \frac{f(T) - f(25)}{f(25)} = -95.27 \times 10^{-6}(T - 25) + 57.5 \times 10^{-9}(T - 25)^2 \quad (1)$$

and can be similarly expressed for mode B. The $R^2$ value of the fits is 0.9998. For mode A, the extracted first- and second-order temperature coefficient of frequency (TCF) are -95.27ppm/°C and 57.5ppb/°C$^2$, respectively. The quadratic fit indicates an expected turnover temperature of 854°C in air. For mode B, the first- and second-order TCF are -95.43ppm/°C and 55.8ppb/°C$^2$, respectively.

The extracted Q and $k_t^2$ of device modes A and B at each temperature are shown in Fig. 5. The quality factor of mode A is fairly stable and exhibits a 33% decrease with temperature, from 415 at 25°C to 274 at 500°C. The Q of mode B is less stable but still decreases 25%, from 508 to 377 over the temperature range, indicating increased damping effects at elevated temperatures. The $k_t^2$ of each single-mode changes during the temperature rising. However, the sum of $k_t^2$ for both modes experiences only a small rise and is fairly stable. This fluctuation of coupling in individual modes can be explained by the differences in TCF between the two modes.

## IV. CONCLUSION

This letter presents a first experiment on the high-temperature (500°C) characteristics of a laterally vibrating (shear horizontal mode) Lithium niobate MEMS resonator array. The LiNbO$_3$ laterally vibrating device displays recoverability of resonant frequency after 500°C air exposure and high and stable electromechanical coupling, $k_t^2$. These results, along with the large frequency variation with temperature reported in this work, support the use of uncompensated LiNbO$_3$ laterally vibrating resonators as low loss, high sensitivity extreme environment sensors (e.g. infrared sensors for space exploration and other high-temp applications).